\documentclass[letterpaper]{article} 
\usepackage{aaai25}  
\usepackage{times}  
\usepackage{helvet}  
\usepackage{courier}  
\usepackage[hyphens]{url}  
\usepackage{graphicx} 
\usepackage{natbib}  
\usepackage{caption} 
\nocopyright
\usepackage{algorithm}
\usepackage{algorithmic}
\usepackage{subcaption}
\usepackage{booktabs}
\usepackage{multirow}
\usepackage{array}
\usepackage{lipsum} 
\usepackage{amsmath}
\usepackage{amssymb}
\usepackage{adjustbox}
\usepackage{xcolor}
\usepackage{newfloat}
\usepackage{listings}
\DeclareCaptionStyle{ruled}{labelfont=normalfont,labelsep=colon,strut=off} 
\floatstyle{ruled}
\newfloat{listing}{tb}{lst}{}
\floatname{listing}{Listing}
\title{StockTime: A Time Series Specialized Large Language Model Architecture for Stock Price Prediction}
\author{
\textbf{Shengkun Wang}$^{1}$, \textbf{Taoran Ji}$^{2}$, \textbf{Linhan Wang}$^1$, \textbf{Yanshen Sun}$^1$,\\
\textbf{Shang-Ching Liu}$^3$, \textbf{Amit Kumar}$^2$, \textbf{Chang-Tien Lu}$^1$
}
\affiliations{
    \textsuperscript{\rm 1}{Virginia Tech},
\textsuperscript{\rm 2}{Texas A\&M University - Corpus Christi},
\textsuperscript{\rm 3}{University of Hamburg}
}

\usepackage{bibentry}

\begin{document}

\maketitle

\begin{abstract}
The stock price prediction task holds a significant role in the financial domain and has been studied for a long time. Recently, large language models (LLMs) have brought new ways to improve these predictions.  While recent financial large language models (FinLLMs) have shown considerable progress in financial NLP tasks compared to smaller pre-trained language models (PLMs), challenges persist in stock price forecasting. Firstly, effectively integrating the modalities of time series data and natural language to fully leverage these capabilities remains complex. Secondly, FinLLMs focus more on analysis and interpretability, which can overlook the essential features of time series data. Moreover, due to the abundance of false and redundant information in financial markets, models often produce less accurate predictions when faced with such input data. In this paper, we introduce StockTime, a novel LLM-based architecture designed specifically for stock price data. Unlike recent FinLLMs, StockTime is specifically designed for stock price time series data. It leverages the natural ability of LLMs to predict the next token by treating stock prices as consecutive tokens, extracting textual information such as stock correlations, statistical trends and timestamps directly from these stock prices. StockTime then integrates both textual and time series data into the embedding space. By fusing this multimodal data, StockTime effectively predicts stock prices across arbitrary look-back periods. Our experiments demonstrate that StockTime outperforms recent LLMs, as it gives more accurate predictions while reducing memory usage and runtime costs.

\end{abstract}

\section{Introduction}
In the financial domain, numerous tasks aim at a common goal: to aid in decision-making by identifying factors that influence market dynamics and achieving arbitrage opportunities in the market.
 Stock price prediction is a crucial task because it directly captures these arbitrage opportunities. For this reason, the application of machine learning methods to predict stock prices has been explored since the last century, underscoring its foundational role in financial domain \cite{kamijo1990stock}. 

In recent years, research on LLMs has rapidly expanded across various domains, including finance. Currently, there is a growing trend to use instruction fine-tuning alongside in-context learning to train FinLLMs, adapting general LLMs for specialized tasks within the financial sector \cite{lee2024survey}, as illustrated in Figure 1. 
\begin{figure}[h]
    \centering
    \includegraphics[width=\columnwidth]{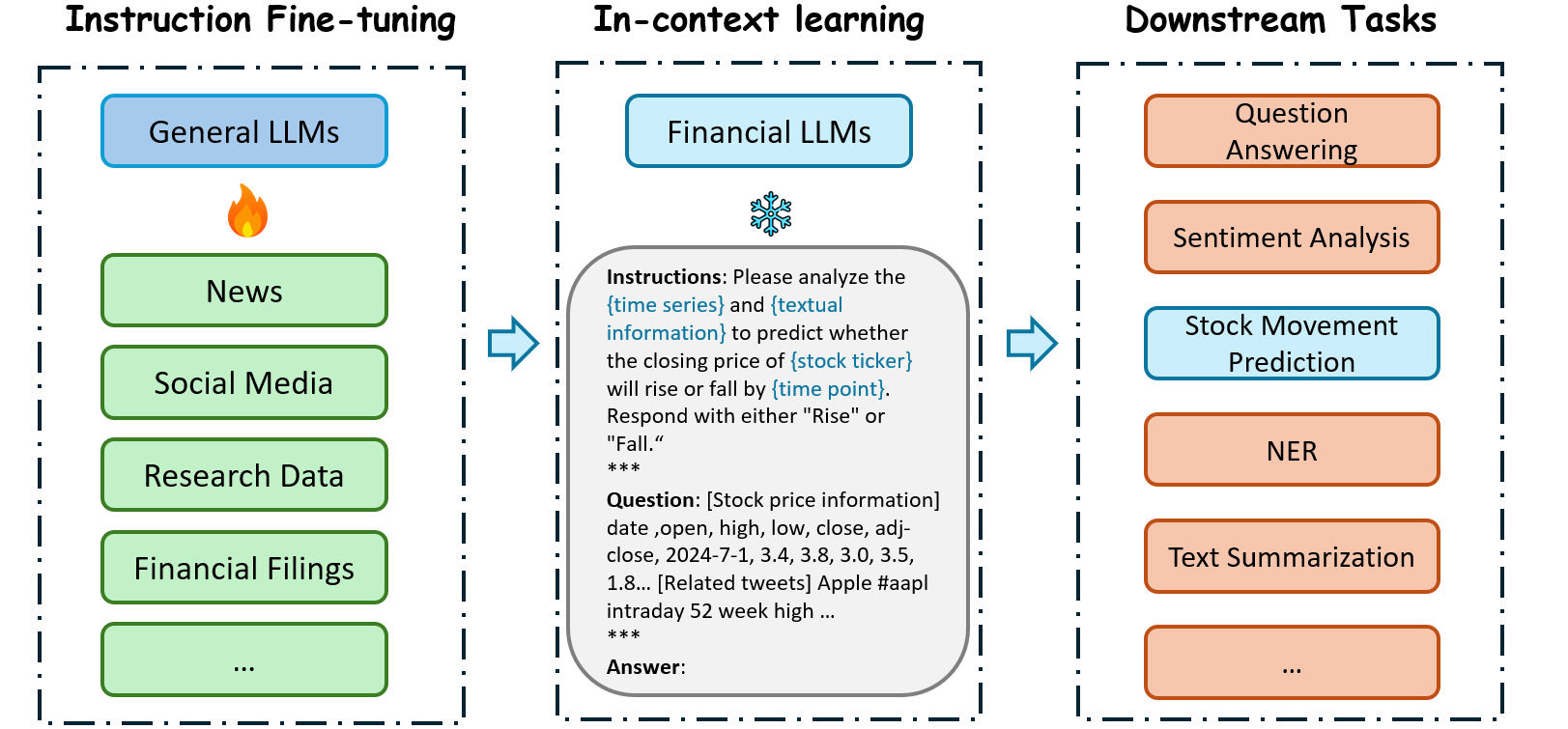}
    \caption{The framework of existing FinLLMs. By applying instruction fine-tuning to general LLMs, FinLLMs update their model parameters. Then, they use different prompts to address various downstream tasks. }
    \label{fig:p1}
\end{figure}
Compared to PLMs, FinLLMs are no longer constrained to specific lookback and prediction 
lengths, which can provide a more comprehensive analysis of historical data and capture long-term trends and patterns in stock market. However, despite their potential, existing FinLLMs primarily focus on interpreting and analyzing publicly available information. In the information-saturated financial markets, these models often struggle to extract the key factors that truly influence stock prices. As a result, FinLLMs tend to underperform compared to smaller autoregressive models when it comes to stock price movements prediction. This is partly because autoregressive models are specifically tailored to model time-dependent data, enabling them to effectively incorporate past information directly into their predictions. Additionally, due to the limited size of autoregressive models, they often perform more efficient input processing and filtering before making predictions.

Although the primary focus of FinLLMs remains on enhancing decision-making and analysis by integrating textual information, the inherent characteristics of LLMs make them versatile tools for a variety of tasks. 
Their capability to handle inputs and outputs of any length and their proficiency in multi-step generation make them particularly suitable for time series prediction tasks. Previous research has demonstrated the viability of using LLMs for such purposes \cite{nie2023timeseriesworth64,jin2023time}. However, time series data cannot be precisely described in discrete natural language, which complicates the direct application of LLMs for understanding time series without aligning detailed textual information. Furthermore, due to the unique characteristics of stock price data, such as sudden fluctuations triggered by unforeseen events and complex correlations between industries and companies, the use of LLMs for financial time series prediction is still in its early stages.

To address the aforementioned problems, we propose an effective LLM-based framework named StockTime, specifically tailored for predicting stock prices using time series data. Initially, we segment and embed stock prices into different patches, then generate textual information including correlations, trend movements and timestamps from these patches. Furthermore, an autoregressive encoder captures the temporal information from the stock prices, which is then fused with the textual information in the latent space of the LLM. By freezing the LLM and training only the integrated embedding and projection layers of the stock time series, we significantly reduce training costs and enable quick adaptation. This method transforms the LLM, typically focused on next-token prediction, into an autoregressive forecaster that is not bound by a specific lookback window. 
Unlike existing financial LLMs, our method does not incorporate any extraneous textual information; it relies solely on the inherent time series data of stock prices.
The mean contributions of this paper are summarized as:

\textbullet\  We present StockTime, an effective framework that leverages the predictive capabilities of LLMs without requiring fine-tuning. It utilizes the LLMs inherent token transitions to extrapolate future stock prices.

\textbullet\ We extract correlation, statistical trends and timestamps from stock prices and seamlessly integrate them with stock time series data in the embedding space, transforming LLMs into an autoregressive forecaster that is not constrained by a specific lookback window.

\textbullet\ We conducted experiments on multi-frequency real-world datasets to validate the design of our proposed method, demonstrating its superiority over existing LLMs.

\section{Related Work}
\subsection{Stock Prediction}
Stock prediction tasks predominantly classify into two main categories: technical analysis and fundamental analysis. The key distinction between them lies in the type of data they utilize; specifically, technical analysis focuses solely on numerical features. Recently, deep learning methods have been extensively employed to enhance stock prediction within the realm of technical analysis. \citet{tang2020add} employed convolutional neural networks to augment training samples by incorporating diverse excess and market features, thus improving prediction performance. Similarly, \citet{sunny2020deep} utilized Long Short-Term Memory (LSTM) networks to capture temporal dependencies in stock prices. Furthermore, \citet{feng2019enhancing} enhanced model robustness against the inherent stochasticity of price variables by integrating adversarial training with perturbations in the feature space. Lastly, \citet{li2024master} developed a transformer-based model that not only models momentary and cross-time stock correlations but also leverages market information for automatic feature selection. With the continuous advancements in NLP technologies, fundamental analysis in stock prediction has increasingly incorporated diverse data sources. Recent studies leverage news \cite{bl2023combined}, social media \cite{wang2023alerta}, and other textual data to predict stock movements. Additionally, there is a growing interest in utilizing visual and auditory modalities for analysis, such as candlestick charts \cite{cagliero2023shortlisting} and earnings calls \cite{wang2024ama}. 
\subsection{Financial LLMs}
In recent years, advancements in LLMs have led to further exploration of fundamental analysis. \citet{wu2023bloomberggpt} introduced BloombergGPT, the first financial LLM with 50 billion parameters. This model was pre-trained on a mixed dataset from both general and financial domains, but it has not been publicly released. Then, the Fingpt team released an open-source model that applies instruction fine-tuning using low-rank adaptation methods and news data to predict stock movements \cite{yang2023fingpt}. Besides, \citet{xie2023pixiu} developed FinMA, which performs multi-task instruction tuning on LLaMA, utilizing a specially constructed dataset. Additionally, Ploutos \cite{tong2024ploutos} utilizes instruction-based methods to enhance financial predictions. These works primarily focus on the textual processing capabilities of LLMs. Diverging from these fundamental analysis approaches that emphasize text, Alpha-GPT \cite{wang2023alpha} introduces a new alpha mining paradigm that focuses on numerical features, yet still requires manual instruction. Our proposed method uniquely considers time series data and derives textual information directly from it, effectively bridging the modality gap that arises when directly combining time series and language tokens.


\subsection{LLMs for Time Series}
Given the impressive performance of LLMs in visual and auditory multimodal capabilities, researchers are exploring their potential in the realm of time series analysis \cite{zhang2024large}. This interest is driven by the desire to extend the versatile applications of LLMs beyond traditional text and media, offering new insights and methodologies for analyzing sequential data. Recognizing the limitations of Byte Pair Encoding (BPE) tokenization, which often breaks single numbers into tokens that do not align with the digits, \citet{gruver2024large} propose a novel tokenization strategy to ensure distinct and consistent tokenization across different floating point numbers.
Additionally, \citet{nie2023timeseriesworth64} introduce a method for segmenting time series into subseries-level patches, which are then used as inputs to the model. \citet{zhou2023one} explore the application of a frozen pre-trained GPT-2 for time series forecasting, where positional embedding layers and self-attention blocks are retained during the finetuning process.
\begin{figure*}[h]
    \centering
    \includegraphics[width=\textwidth]{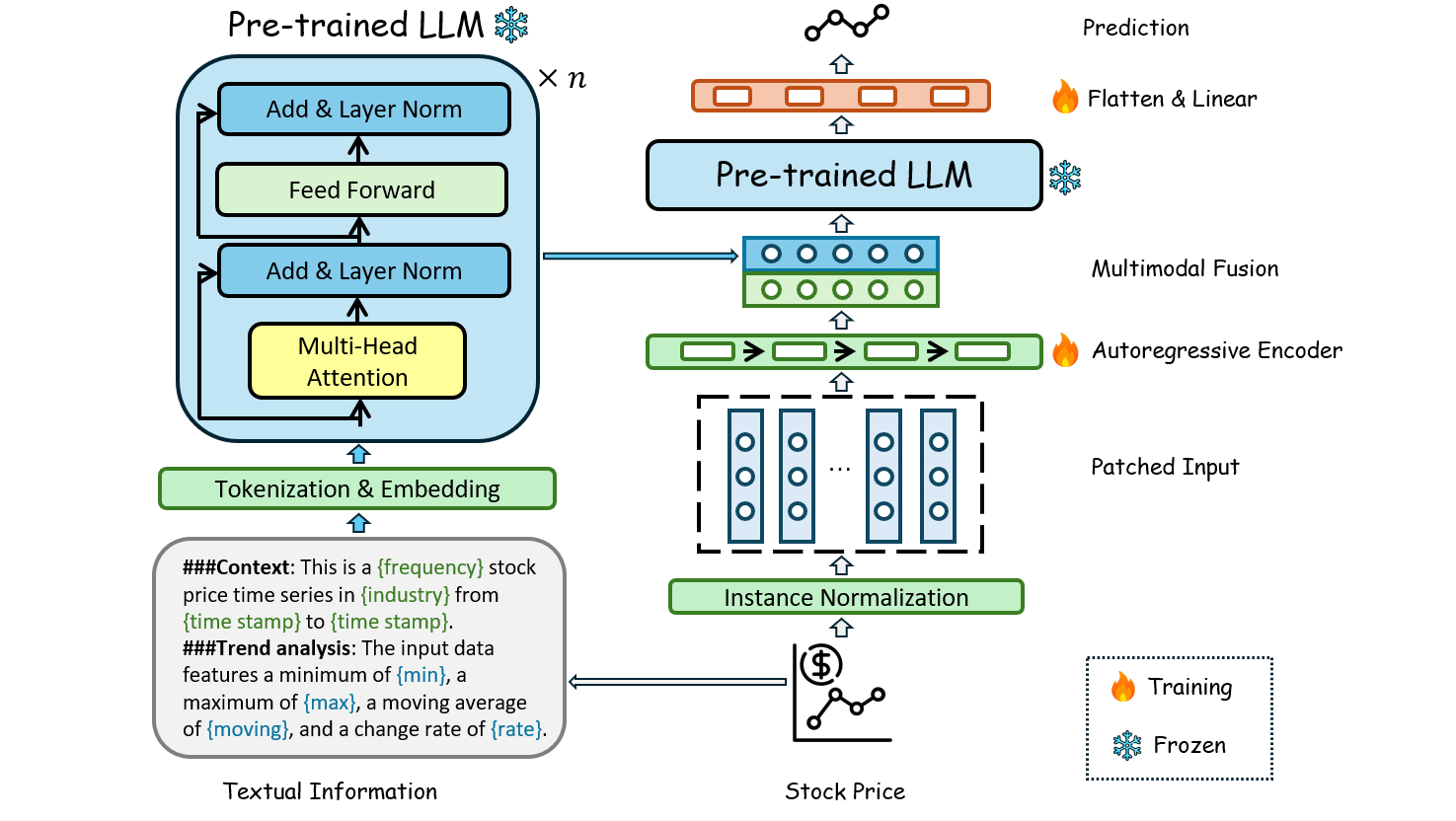}
    \caption{The StockTime framework operates as follows: (1) Stock correlations, statistical trends, and time step information are extracted from stock prices and processed as textual information through a frozen LLM. (2) Stock time series data is segmented and embedded, then passed through an autoregressive encoder to be integrated with textual information, which is subsequently processed by a pre-trained LLM. (3) After learning the multimodal information, the off the-shelf LLM as an autoregressive forecaster to predict the next token, which corresponds to the predicted stock price.}
    \label{fig:p2}
\end{figure*}
\citet{xue2023promptcast} proposes a prompt-based approach to time series forecasting by converting numerical time series into text prompts and employing a sentence-to-sentence forecasting methodology. Time-LLM \cite{jin2023time} reprograms time series data into text prototypes, leveraging the LLaMA-7B model for processing.
\citet{rasul2023lag} builds a univariate probabilistic time series forecasting model based on the LLaMA architecture, enhancing model accuracy and applicability. Furthermore, \citet{liu2024autotimes} formulate time series as prompts, extending the contextual window for prediction and introducing an in-context forecasting method.

\section{Methodology}
\subsection{Problem Definition}
Given a stock price $p$ within a pre-selected stock dataset $P \in \mathbb{R}^{S \times D}$, where $D$ denotes the number of days and $S$ represents the number of stocks. With a lookback window of $d$ days, stock $s$ price is ${p}_{s,1:d} = \{{p}_{s,1}, \ldots, {p}_{s,d}\} \in \mathbb{R}^{1 \times d}$, we aim to forecast the stock price for the subsequent $x$ days, ${p}_{s,d+1:d+x} = \{{p}_{s,d+1}, \ldots, {p}_{s,d+x}\} \in \mathbb{R}^{1 \times x}$. Additionally, the textual information derived from the stock price is integrated with the stock price data in the latent space at time $t$.
This study relies exclusively on stock price data as input, which defines it as a univariate stock price prediction task. The goal is to train the LLM-based model $f(\cdot)$ to predict the future stock price  $\hat{p}$ for a forecast period of $x$ days based on a lookback period of $d$ days. The process can be described as: $f({p}_{s,1:d}) \rightarrow \hat{{p}}_{s,d+1:d+x}$.

\begin{table*}[!htbp] 
    \centering
    \begin{tabular}{lllllc} 
        \toprule
        \textbf{Dataset} & \textbf{Date} & \textbf{Frequency} & \textbf{Time Steps} & \textbf{Modalities} & \textbf{Number of Stocks}
 \\
        \midrule
        S\&P 100-H & 2023-06-30 9:30 to 2024-07-16 15:30 & Hourly & 1822  & time series & 100\\
        S\&P 100-D & 2014-06-30 to 2024-06-28 & Daily & 2518& time series &  97 \\
        Bigdata23 & 2020-06-01 to 2023-05-31 & Daily & 756 & time series, text & 42 \\
        Bigdata22 & 2019-07-05 to 2020-06-30 & Daily & 362 & time series, text & 50 \\
        ACL18 & 2014-01-02 to 2015-12-30 & Daily & 696 & time series, text & 87 \\
        CIKM 18 & 2017-01-03 to 2017-12-28 & Daily & 231 & time series, text & 47 \\
        \bottomrule
    \end{tabular}
    \caption{Overview of datasets.}
    \label{tab:dataset_overview}
\end{table*}

\subsection{Stocktime Overview}
The Stocktime architecture is illustrated in Figure 2. Our method consists of four main components: (1) patched input, (2) autoregressive encoder, (3) multimodal fusion, and (4) token-level prediction.
Initially, we process stock correlations and statistical information with timestamps through a frozen LLM. Then, we feed the patched stock price through the autoregressive encoder and concatenate the preprocessed information in the embedding space. The fused input is subsequently passed through a frozen LLM to obtain the output representations. Finally, these representations are flattened and linearly projected to derive the final forecasts. 
In the following sections, we will explain the function of each component in detail. Unlike traditional financial LLMs, which typically require textual instructions and fine-tuning of the backbone model, StockTime is directly optimized using only stock price data and a few training epochs. Our framework ensures high efficiency and significantly reduces resource requirements compared to building FinLLMs from scratch or fine-tuning existing general LLMs.

\subsection{Patched input}
Historical stock prices have proven to be strong indicators of future stock trends and are widely referenced in financial literature \cite{fan2024stockmixer}. To effectively capture correlations, we first normalized each stock price to have a mean of zero and a standard deviation of one using reversible instance normalization. Next, we segmented the stock prices into consecutive, non-overlapping patches, implicitly capturing the correlations between stocks through shared parameters. A stock price time series can be divided into $n$ patches, with the $i$-th patch $\mathbf{h}_i$ of length $l$ defined as:

\begin{equation}
\mathbf{h}_i = \{p_{s,(i-1)l+1}, \ldots, p_{s,il}\}, \quad i \in \{1, \ldots, n\},
\end{equation}
Each patch is treated as a basic token to form a compact sequence of input tokens, thereby reducing computational burdens. However, time series data cannot be directly edited or described losslessly in natural language, posing significant challenges in directly adapting LLMs to understand time series without resource-intensive fine-tuning. To address this issue, we developed a textual template that includes stock correlations, statistical trends, and timestamp information, all derived from stock time series data. This textual information is then fused with the corresponding patched stock price tokens, as detailed in the multimodal fusion section.


\subsection{Autoregressive Encoder}
LLMs typically exhibit reduced sensitivity when processing high-precision numerals without external information, presenting substantial challenges in accurately addressing practical forecasting tasks over long horizons. While recurrent neural networks (RNNs) are preferred for sequential data processing due to their intrinsic ability to manage sequential dependencies, they tend to struggle with long-term dependencies and suffer from issues such as vanishing gradients, which limit their effectiveness in processing extended sequences. In contrast, LSTM networks, with their specialized gating mechanisms are better suited for handling long-range dependencies in time series data. Therefore, we adopted an LSTM layer as part of the encoder to effectively encode the patched stock price data. At each time step for stock price, the recurrent unit learns hidden representations by jointly considering the input $\mathbf{h_i}$ and the previous hidden state to capture the sequential dependencies. By adding a fully connected layer after the LSTM layer, the $\textbf{Autoregressive Encoder}(\cdot)$ projects the sequential dependencies of the stock price segments from dimension $l$ into the LLM's model dimension $d_{\text{llm}}$ in the latent space as price embedding:

\begin{equation}
    \mathbf{pe}_i = \textbf{Autoregressive Encoder}(\mathbf{h_i}), \mathbf{pe}_i \in \mathbb{R}^{1 \times d_{\text{llm}}} .
\end{equation}

\subsection{Multimodal Fusion}
Throughout the previous operation, we obtained the stock price embedding. To integrate temporal sequences with textual information that LLMs can understand, and to enable the model to comprehend correlations among different stocks, we constructed a textual template that includes various details corresponding to each stock price patch. This template comprises three key components: 1) the time series frequency of stock prices across different datasets, 2) the industry classification of the various stocks, and 3) the statistical details including minimum, maximum, and average values, along with the average rate of change and the corresponding timestamps for the stock price patches. All of this information is derived directly from the stock price data itself, as illustrated in Figure 2. We then tokenized and embedded the textual input, passing it through an  off the-shelf LLM to transform the information into the embedding space to have the textual embedding $\mathbf{ce_i}$:

\begin{equation}
    \mathbf{ce}_i = \textbf{LLM}(c_i), \mathbf{ce_i} \in \mathbb{R}^{1 \times d_{\text{llm}}} .
\end{equation}

Through our experiments, we discovered that aligning stock price data with textual information cues in StockTime leads to a significant improvement in prediction outcomes. This finding suggests that explicitly incorporating stock correlations into the textual information yields better results than merely capturing stock correlations implicitly through shared parameters. The textual embedding $\mathbf{ce}_i$ is processed separately by a frozen LLM and then concatenated with $\mathbf{pe}_i$ in the latent space. This approach allows the textual embedding to be integrated with the corresponding price patch embedding without increasing the context length. The procedure is as follows:

\begin{equation}
    \mathbf{e}_i = \mathbf{pe}_i + \mathbf{ce}_i,\mathbf{e}_i \in \mathbb{R}^{1 \times d_{\text{llm}}} .
\end{equation}



\begin{table*}[htbp]
\centering
\begin{tabular}{lcccccccc}
\toprule
\textbf{Data} & \multicolumn{2}{c}{BigData23} & \multicolumn{2}{c}{BigData22} & \multicolumn{2}{c}{ACL18} & \multicolumn{2}{c}{CIKM18} \\
\cmidrule(lr){1-3} \cmidrule(lr){3-5} \cmidrule(lr){5-7} \cmidrule(lr){7-9}
 \textbf{Method} & ACC. & MCC & ACC. & MCC & ACC. & MCC & ACC. & MCC \\
\midrule
Mathstral-7B & 0.497 & 0.003 & 0.507 & -0.027 & 0.486 & 0.005 & 0.502 & 0.031 \\
LLaMA3-8B & 0.511 & 0.016 & 0.502 & 0.024 & 0.519 & 0.047 & 0.495 & 0.008 \\
GPT-4o mini & 0.518 & \textbf{0.076} & \textbf{0.521} & 0.036 & 0.525 & 0.057 & 0.513 & 0.023 \\
FinMA & 0.506 & 0.041 & 0.505 & 0.013 & 0.512 & 0.026 & 0.494 & \textbf{0.074} \\
StockTime & \textbf{0.524} & 0.061 & 0.515 & \textbf{0.041} & \textbf{0.539} & \textbf{0.062} & \textbf{0.517} & 0.069 \\
\bottomrule
\end{tabular}
\caption{Experiments on four stock price and tweets datasets, with the best results highlighted in bold. Comparison methods include General LLMs and FinLLMs}
\label{tab:performance_metrics_transposed}
\end{table*}
\subsection{Prediction}

Since LLMs are primarily trained on discrete textual data, which differs from the continuous numerical nature of stock prices, we exploit the LLMs' capability to predict the next token based on preceding tokens to achieve predictions of arbitrary lengths. As previously mentioned, we divide the historical stock price embeddings into $n$ consecutive patches, with each patch having a length of $l$. The token embeddings $\mathbf{e}_i$ are fed into the off-the-shelf LLM and then projected back to the prediction patch $\mathbf{\hat{h}}_i$. The training objective is to independently generate the next tokens $\{\mathbf{\hat{h}}_2, \ldots, \mathbf{\hat{h}}_{n+1}\}$. Each predicted patch is supervised by the token-wise ground truth to optimize the parameters of the embedding and projection layers, which are implemented as simple linear layers. The loss function used is Mean Squared Error (MSE):

\begin{equation}
    \text{MSE} = \frac{1}{nl} \sum_{i=2}^{n} \|\mathbf{\hat{h}}_i - \mathbf{h}_i\|_2^2.
\end{equation}




\section{Experiments}
In this section, we conduct experiments to answer the following four research questions:
\begin{itemize}
    \item \textbf{Q1}: How does the performance of StockTime compare with general LLMs and FinLLMs on the datasets that have stock price and tweets?
    \item \textbf{Q2}: Will FinLLMs that have been fine-tuned on extensive textual data perform better in stock price prediction?
    \item \textbf{Q3}: Is the proposed LLM architecture more effective for stock price forecasting compared to other LLM-based time series methods?

    \item \textbf{Q4}: How do the individual model components and hyper-parameters impact the performance of StockTime?

\end{itemize}

\subsection{Experimental Setup}

\noindent \textbf{Datasets.} According to S\&P\footnote{\url{https://www.spglobal.com}}, U.S. stocks are categorized into 11 sectors: \textit{Information Technology}, \textit{Financials}, \textit{Health Care}, \textit{Energy}, \textit{Industrials}, \textit{Consumer Discretionary}, \textit{Consumer Staples}, \textit{Utilities}, \textit{Communication Services}, \textit{Materials}, and \textit{Real Estate}. To ensure the datasets accurately represent the stock market, we sourced historical stock price data for S\&P 100 companies from Yahoo Finance\footnote{https://finance.yahoo.com} for the period from June 30, 2014, to June 28, 2024. We excluded three companies due to insufficient historical data length. The data for the remaining companies is distributed across the aforementioned S\&P sectors.
Since our framework does not require textual data or analysis, the training and inference time is significantly reduced. Consequently, we also created a hourly medium-frequency  stock dataset using companies from the S\&P 100, covering the period from June 30, 2023, 9:30  to July 16, 2024, 15:30. This dataset was used to evaluate StockTime's performance in hourly medium-frequency trading scenarios.
Additionally, we adopt four datasets with textual data aligned with stock time series data: Bigdata23 \cite{wang2023stock}, Bigdata22 \cite{soun2022accurate}, ACL18 \cite{xu2018stock}, and CIKM18 \cite{wu2018hybrid}. For these four datasets, our experiments with FinLLMs and general LLMs incorporated stock price and textual data, while for Stocktime, we only used the adjusted close price for experiments. All dataset statistics are presented in Table 1.

\vspace{1mm}

\noindent \textbf{Implementation Details.}
All the experiments are conducted using PyTorch \cite{paszke2019pytorch} on NVIDIA A100 GPUs. We employ the Adam optimizer \cite{kingma2014adam} with an initial learning rate $1e -3 $ and and we selected the best hyperparameters based on the IC performance in the validation stage. The lookback window is choosen from \{16,32,64,128,256\} and the batch size is chosen from \{16, 32, 64\}. We set the number of training epochs as 10. Unless otherwise specified, we use LLaMA3-8B\footnote{https://huggingface.co/meta-llama/Meta-Llama-3-8B} as the default base LLM and use MSE loss for model optimization. Each experiment was repeated 3 times and the average performance was reported.

\noindent \textbf{Baselines.} We compare the performance of our framework with several LLMs specifically designed for stock movement prediction and time series methods used for stock price prediction. For the selection of baseline models, we focus on those that are open-source or have accessible APIs, allowing us to conduct thorough testing. The baselines include:

\begin{itemize}
    \item LLMs for Time Series Models:
    \begin{itemize}
        \item FPT \cite{zhou2023one}: A model uses LLMs, with GPT-2 as the backbone, to extract sequential patterns from time series data.
        \item Times-LLM \cite{jin2023time}: This model reprograms the input time series into text-based prototype representations, making them more naturally suited to language models' capabilities.
        \item AutoTimes \cite{liu2024autotimes}: This model use in-context forecasting approach that formulates time series as prompts, and a timestamps as position embeddings.
    \end{itemize}
    
    \item Financial LLM:
    \begin{itemize}
        \item FinMA \cite{xie2023pixiu}: An open-source FinLLM based on LLaMA, trained using instruction fine-tuning techniques.
    \end{itemize}

\begin{table}[htbp]
\centering
\begin{tabular}{lcccc}
\toprule
\textbf{Data} & \multicolumn{2}{c}{S\&P 100-D} & \multicolumn{2}{c}{S\&P 100-H} \\
\cmidrule(lr){0-1}\cmidrule(lr){2-3} \cmidrule(lr){3-5}
\textbf{Method} & MSE & IC & MSE & IC \\
\midrule
Times-LLM & 0.167 & 0.007 & 0.194 & 0.011 \\
AutoTimes & 0.179 & 0.012 & 0.183 & 0.009 \\
FPT & 0.182 & 0.003 & 0.205 & 0.006 \\
\textbf{StockTime} & \textbf{0.146} & \textbf{0.018} & \textbf{0.178} & \textbf{0.014} \\
\bottomrule
\end{tabular}
\caption{Experiments on S\&P 100 intraday and hourly medium-frequency datasets are presented. Comparison methods include LLMs for time series modeling.}
\label{tab:2}
\end{table}

    \item General LLMs:
    \begin{itemize}
        \item  Mathstral-7B \cite{jiang2023mistral}, LLaMA3-8B \cite{dubey2024llama}, GPT-4o Mini\footnote{https://openai.com/index/gpt-4o-mini-advancing-cost-efficient-intelligence/}: The parameter sizes of these general LLMs are similar to the other baseline models.
    \end{itemize}
\end{itemize}

\noindent \textbf{Metrics.}
Although our primary task is to predict stock prices, the outcomes from financial language models are typically reported as stock price movements, either upward or downward. To fairly evaluate our framework, we adopt four metrics commonly used in stock prediction tasks. For datasets accompanied by textual data, we use accuracy (ACC.), which measures the percentage of correct movement predictions, and Matthews correlation coefficient (MCC), a balanced performance measure for binary classification tasks. For datasets sourced without textual data, we use mean squared error (MSE) quantifies the average squared difference between predicted and actual stock prices, while information coefficient (IC) assesses the rank correlation between predicted changes and actual outcomes.

\subsection{Overall Performance and Analysis}
The comparison to FinLLM and general LLMs is presented in Table 2, while the comparison between StockTime and the recent methods for LLMs for time series model is shown in Table 3. Most of the baselines’ results on the benchmarks are reported using their original settings and all of them adopt the same optimization loss in ensuring fair.
We address the first three research questions by analyzing the experimental results:

1) Compared to FinLLM, our framework outperformed them on most datasets containing textual data, achieving up to a 5\% improvement in stock price movement prediction. This demonstrates that our approach not only saves resources and time by eliminating the need for fine-tuning but also maintains high accuracy. Moreover, FinLLM did not show a significant advantage over general LLMs, indicating that even after extensive fine-tuning with financial data, the improvement in stock price prediction remains limited. This suggests that future efforts in using FinLLM for stock-related tasks should focus more on the processing of textual data and the intrinsic characteristics of time series data.
\begin{table}[htbp]
\centering
\begin{tabular}{lcccc}
\toprule
\textbf{Data} & \multicolumn{2}{c}{S\&P 100-D} & \multicolumn{2}{c}{S\&P 100-H} \\
\cmidrule(lr){0-1}\cmidrule(lr){2-3} \cmidrule(lr){3-5}
\textbf{Method} & MSE & IC & MSE & IC \\
\midrule
RNN & 0.326 & -0.014 & 0.315 & 0.005 \\
LSTM & 0.304 & 0.006 & 0.288 & -0.017 \\
ALSTM & 0.271 & -0.008 & 0.292 & 0.003 \\
\textbf{StockTime} & \textbf{0.146} & \textbf{0.018} & \textbf{0.178} & \textbf{0.014} \\
\bottomrule
\end{tabular}
\caption{Experiments are conducted on S\&P 100 intraday and hourly medium-frequency datasets to compare the performance of StockTime with autoregressive models.}
\label{tab:2}
\end{table}
2) While general LLMs have the advantage of not requiring textual information conversion and preprocessing, their performance in stock price prediction was suboptimal compared to StockTime that solely use stock time series data. Although StockNet incorporates a frozen LLM, it treats time series as token outputs. This design enables the model to better understand continuous time series data, even though it was originally designed to generate discrete text.
Additionally, we argue that the performance of LLMs is hindered by the low quality of current stock-related textual data, which is frequently affected by misinformation and excessive redundancy. Moreover, due to stock prices are highly sensitive to market information, the price movements themselves capture a substantial portion of the underlying market sentiment. This makes it more reasonable to use LLM  architectures focused  on analyzing stock time series data.

3) By focusing solely on time series data, we have made it feasible to employ LLM architectures for hourly and medium-frequency trading. Compared to other LLM-based time series methods, StockTime outperforms all baseline approaches on intraday and hourly medium-frequency datasets, demonstrating that autoregressive methods are more effective at capturing temporal information. Additionally, given the unique nature of the stock market, where correlations between stocks and statistical trends are critical, StockTime’s approach seamlessly integrates textual information with stock time series data, effectively addressing these factors and leading to superior performance.

\begin{figure*}[!htb]
  \centering
  \begin{subfigure}{.5\textwidth}
    \centering
    \includegraphics[width=\linewidth]{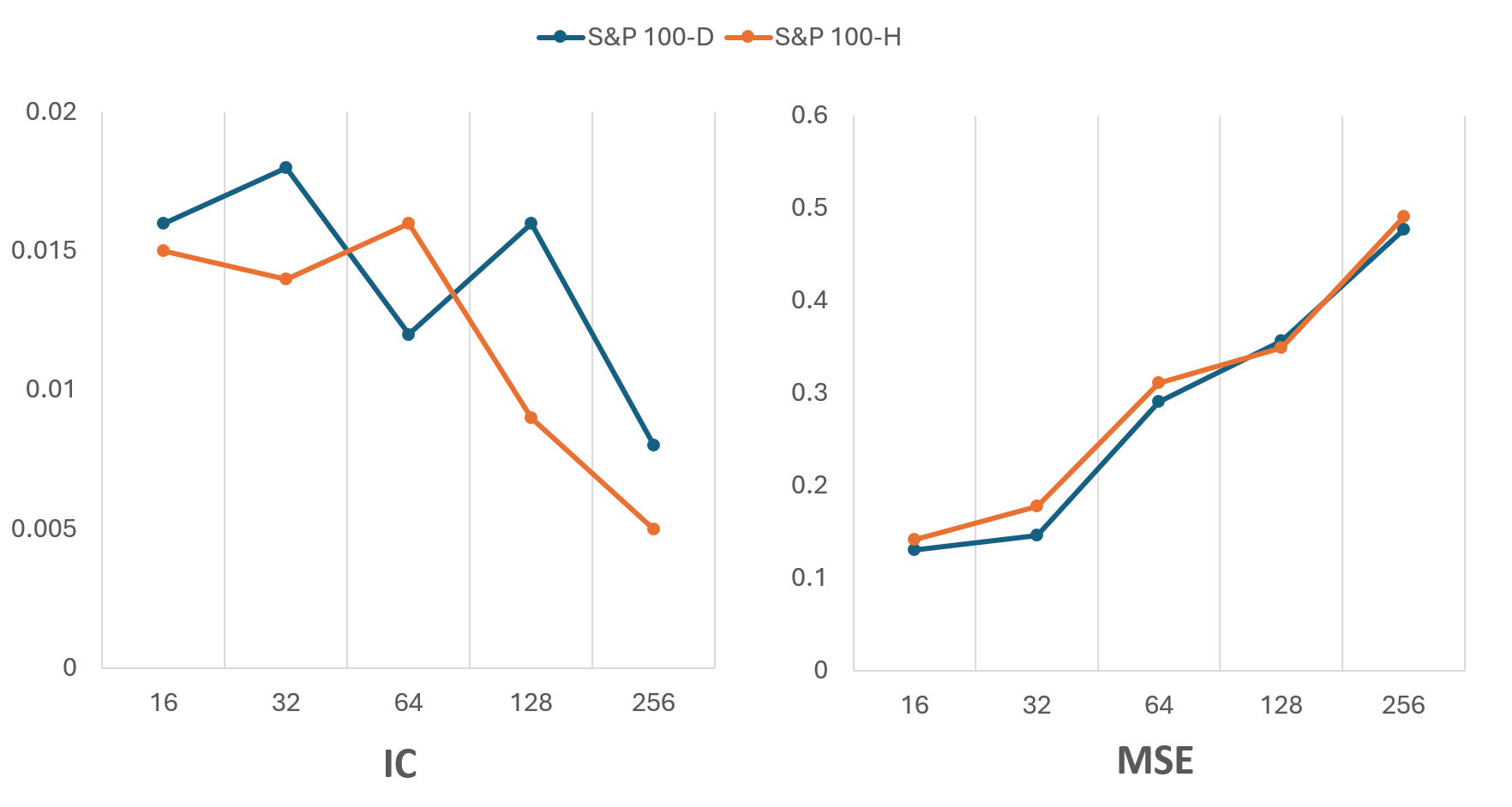}
    \caption{\small Lookback window length}
    \label{fig:sub1}
  \end{subfigure}%
  \begin{subfigure}{.5\textwidth}
    \centering
    \includegraphics[width=\linewidth]{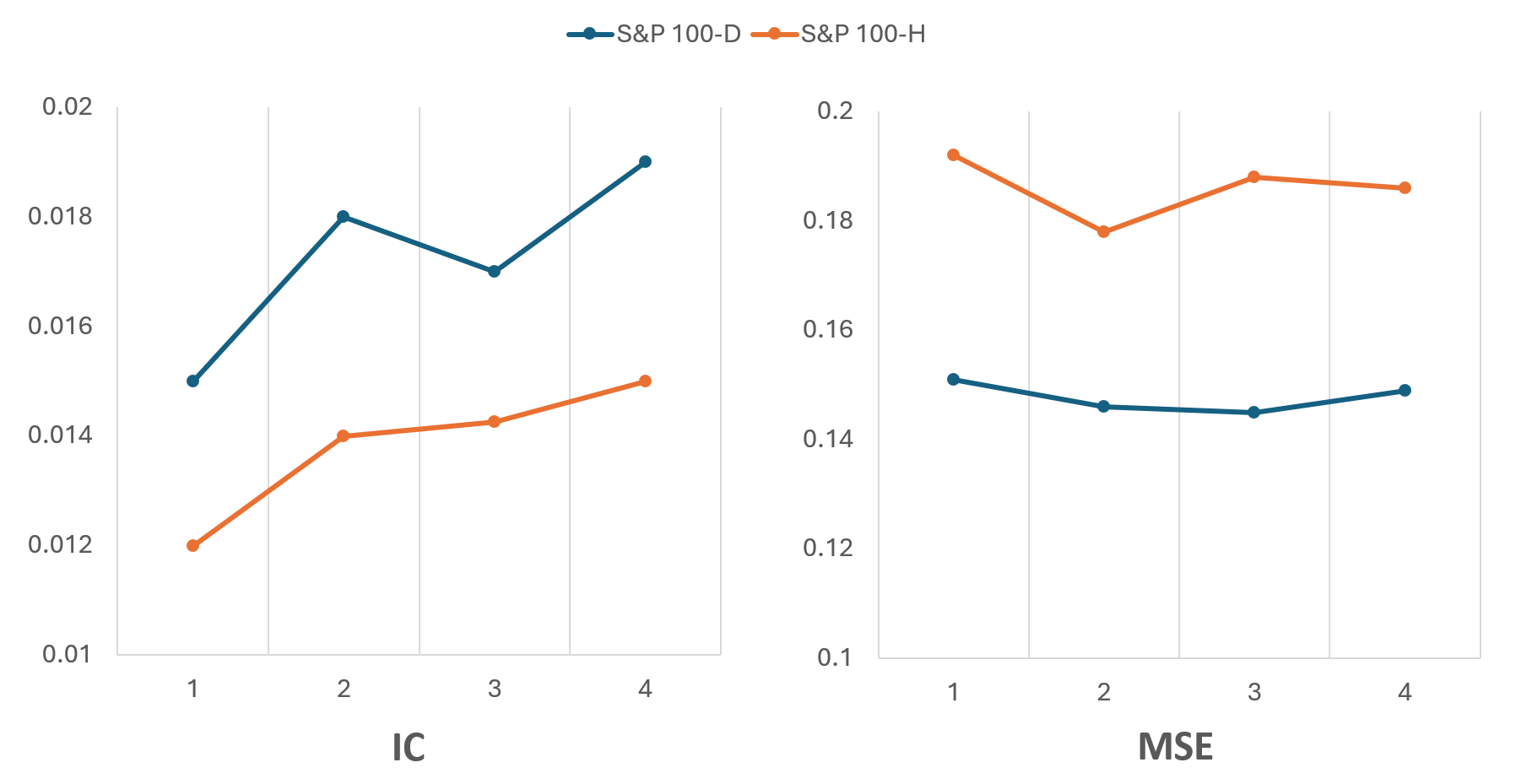}
    \caption{Autoagressive encoder layer}
    \label{fig:sub2}
  \end{subfigure}
  \caption{Hyperparameter sensitivity analysis of different lookback window lengths and autoregressive encoder layers.}
  \label{fig:combined}
\end{figure*}

\subsection{Autoregressive Models Comparsion}
StockTime, a framework that leverages the LLM architecture for stock price prediction, offers faster and more effective performance compared to larger LLMs. However, the advantages of using a large model architecture might be questioned if smaller autoregressive models can achieve similar results. To address this concern, tests were conducted on S\&P datasets using several commonly used autoregressive models for stock price prediction, including 1) RNN, 2) LSTM, and 3) Attention LSTM. The comparison, as shown in Table 4, revealed that StockTime outperformed the autoregressive models in both MSE and IC metrics. These results further validate that utilizing the LLM architecture provides significant benefits for predicting stock time series.

\begin{table}[htbp]
\centering
\begin{tabular}{lcccc}
\toprule
\textbf{Data} & \multicolumn{2}{c}{S\&P 100-D} & \multicolumn{2}{c}{S\&P 100-H} \\
\cmidrule(lr){0-1}\cmidrule(lr){2-3} \cmidrule(lr){3-5}
\textbf{Model Component} & MSE & IC & MSE & IC \\
\midrule
MLP Encoder & 0.161 & 0.012 & 0.191 & 0.008 \\
w.o. Encoder & 0.170 & 0.006 & 0.189 & 0.004 \\
w.o. Fusion & 0.193 & 0.003 & 0.196 & 0.012 \\
GPT2-Backbone & 0.186 & 0.010 & 0.185 & 0.007 \\
\textbf{StockTime} & \textbf{0.146} & \textbf{0.018} & \textbf{0.178} & \textbf{0.014} \\
\bottomrule
\end{tabular}
\caption{Ablation study on autoregressive encoder, multimodal fusion, and backbone LLM conducted on the S\&P 100 datasets.}
\label{tab:2}
\end{table}

\subsection{Ablation Study} 
We answer the fourth research question through ablation study and hyperparameter sensitivity:

\vspace{1mm}

\noindent \textbf{Model Component.}
In this section, we analyze the effectiveness of StockTime by breaking down its individual components. Specifically, we replaced the stock price encoder and the backbone model of the framework, conducting tests on the S\&P 100 datasets. As shown in Table 5, each model component contributed to the overall performance.
For the encoder tests, we substituted the autoregressive encoder with an MLP and a linear layer. The results show that adding an encoder improved the model's performance, and the autoregressive encoder was particularly effective in capturing sequential dependencies compared to the MLP.
When the textual information was removed, StockTime's performance slightly declined, underscoring the importance of integrating multimodal data. For stock price prediction, the fusion of statistical trends and stock correlations led to more accurate predictions, highlighting the necessity of using textual information as hints in LLMs.
In tests with different backbone models, we found that GPT-2 performed slightly worse than LLaMA3. This difference may be attributed to the distinct tokenization methods used by the two models for handling numerical data.

\subsection{Hyperparameter Sensitivity.} 

\vspace{1mm}

\noindent \textbf{Lookback window length.}
We analyze StockTime's stock prediction performance with varying lookback window lengths, as shown in Figure 3a. Based on the model's performance on the S\&P datasets, evaluated using the IC and MSE metrics, a lookback window length of around 32 appears to be optimal. Both shorter and longer window lengths result in a decline in IC and MSE performance.

\vspace{1mm}
\noindent \textbf{Autoagressive encoder layer.}
We analyzed the impact of the number of LSTM layers in the autoregressive encoder on the model's performance in Figure 3b. With the LSTM layer dimension fixed at 256, we observed that the IC metric showed some variation with changes in the number of LSTM layers, while the MSE metric remained unaffected by increasing the number of layers. To enhance model efficiency, we selected two LSTM layers as the optimal configuration.

\section{Conclusion}
In this paper, we proposed StockTime, an efficient LLM-based architecture for stock price prediction. StockTime leverages the inherent token transitions of LLMs to extrapolate future stock prices. Furthermore, it extracts correlations between stocks, statistical trends, and timestamps from stock price data, transforming them into textual information to help LLMs better understand stock time series. This paper demonstrates the potential of efficiently adapting off-the-shelf LLMs for stock price prediction by leveraging only stock price data, rather than fine-tuning on large amounts of textual data. Experiments reveal that the StockTime framework outperforms existing FinLLM and general LLM baselines, suggesting a new direction for LLMs in intraday and hourly medium-frequency stock price prediction.

\bibliography{aaai25}

\end{document}